\documentclass[prb,twocolumn,amsmath,amsfonts,10pt]{revtex4-1}
\usepackage{graphicx}
\usepackage{amsmath}
\usepackage[utf8]{inputenc}
\usepackage{color}
\usepackage{bbold} 

 \definecolor{darkgreen}{rgb}{0,0.5,0} 
 \usepackage{color}
 \usepackage[normalem]{ulem}
 \usepackage{cancel}
 \newcommand{\tdel} [1]{{\color{red}\sout{#1}}}         
\newcommand{\ket}[1]{| \, #1 \, \rangle} 
\newcommand{\bra}[1]{\langle \, #1 \, |} 
\newcommand{\scal}[2]{\bra{#1} \, #2 \, \rangle} 
\newcommand{\myonlinecite}[1]{\hspace{-1 ex} \nocite{#1}\citenum{#1}} 

\begin{document}

\title{Tree tensor network state study of the ionic-neutral curve crossing of LiF}

\author{V. Murg$^{1}$}

\author{F. Verstraete$^{1}$}

\author{R. Schneider$^{2}$}

\author{P. R. Nagy$^{3}$}

\author{\"O. Legeza$^{4}$}

\affiliation{$^1$Fakult\"at f\"ur Physik, Universit\"at Wien, Boltzmanngasse 3, A-1090 Vienna, Austria\\
$^2$Technische Universit\"at Berlin Fakult\"at II - Mathematik und Naturwissenschaften\\
Institut f\"ur Mathematik, Strasse des 17. Juni 136, Berlin, Germany\\
$^3$\tdel{}E\"otv\"os Lor\'and  University, Budapest, P\'azm\'any P. s\'et\'any 1/A, H-1117, Hungary\\
$^4$Strongly correlated systems "Lend\"ulet" research group,\\ Wigner Research Centre for Physics, P.O.Box 49 Hungary
}

\date\today

\begin{abstract}
We present a tree-tensor-network-state (TTNS) method study of the ionic-neutral curve crossing of LiF.
For this ansatz, the long-range correlation deviates from the
mean-field value polynomially with distance, thus for quantum chemical applications
the computational cost could be significantly smaller than that of previous attempts using the
density matrix renormalization group (DMRG) method.
Optimization of the tensor network topology and localization of the avoided crossing are discussed
in terms of entanglement. 
\end{abstract}

\maketitle

\section{Introduction}
\label{sec:intro}

It has been more than a decade ago that the quantum chemistry version 
of the density matrix renormalization group (QC-DMRG) method \cite{white,white-qc} 
has been applied to study the ionic-neutral curve crossing of LiF in order to 
demonstrate that it provides a globally accurate description of the system 
even if the wavefunction changes dramatically transversing the avoided crossing\cite{dmrg_lif}.
In the following years, various theoretical studies have been devoted to investigate 
dissociation curves in diatomic molecules using 
QC-DMRG \cite{chan-pes,reiher-pes,Kurashige2009,boguslawski2013a} and
by now the method has become a rival to the conventional multiconfiguration 
wave function approaches \cite{legeza-rev,reiher-rev,chan-rev}. 
Inclusion of the concepts of entanglement from quantum information
theory (QIT) \cite{legeza03c,legeza04a,rissler06,barcza2010a} has paved the road for identifying 
highly correlated molecular orbitals leading to an efficient construction of active 
spaces \cite{legeza03c,boguslawski2012b} and for characterizing the various types
of correlation effects relevant for chemical bonding \cite{boguslawski2013a,yanai2013}.

In the mean time, a reformulation of DMRG in terms of so-called matrix product 
states (MPS) \cite{ostlund,verstraeteciracmurg08,schollwock2011,Haegeman2013}
has shown that it is only one special case in a much more general set of methods: 
the so-called tensor network states (TNS) 
\cite{verstraetecirac04,murgverstraete07,murgverstraete08,verstraeteciracmurg08,vidal06,changlani09,marti10,marti-rev,schneider-rev,schneider-rev2013,hackbusch,orus-rev}, 
which in certain cases is expected to even outperform QC-DMRG in the near future \cite{murg2010-tree,chan2013-tree}.
A special form of TNS, the tree tensor network states (TTNS) approach \cite{shi2006,Tagliacozzo2007,Corboz2009,Corboz2010} 
was first
applied in quantum chemistry by some of us \cite{murg2010-tree} to present the 
underlying theoretical background and scaling properties of the QC-TTNS algorithm 
while an efficient extension of the QC-DMRG using the tree-like topology 
has been applied recently to dendrimers \cite{chan2013-tree}.
The TTNS approach also plays a fundamental role in {\em hierarchical tensor
decompositions}, recently developed for tensor product approximation
\cite{hackbusch}, see e.g. \cite{schneider-rev,schneider-rev2013}.

Unlike to models with translational symmetry studied usually in condensed matter physics, 
the orbital entanglement is non-constant in the quantum chemical applications. 
Therefore, the optimal arrangement of matrices in MPS-based approaches has 
a tremendous effect on their performance and on the required computational resources 
to reach a given accuracy \cite{legeza2003a,legeza03c,barcza2010a}. 
Similar optimization strategies to find the best tensor topology are also crucial 
in case of the TTNS algorithm \cite{barcza2010a,chan2013-tree}.

In this paper, we discuss the most general version of the QC-TTNS algorithm in which 
the local properties of the tensors can be different for each orbital.  
By studying the ionic-neutral potential curve crossing of the LiF we present an 
optimization strategy to set up tensor topologies which reflect the structure 
of the entanglement bonds between the molecular orbitals as the bond length 
between the Li and F is stretched. We also compare the MPS(DMRG) and TTNS 
convergence properties by changing the order of the tensors only. 
In our study, we calculate the two lowest $^1\Sigma^+$ states of LiF and the corresponding 
one- and two-orbital entropy functions\cite{rissler06,boguslawski2013a}. 
The localization of the avoided crossing in terms of orbital entropy is also discussed.

The setup of the paper is as follows. In Sec.~II we briefly describe the main steps
of the QC-TTNS algorithm and the details of the numerical procedure used to determine 
the optimal tensor arrangements. Sec.~III contains the numerical results and analysis of the observed
trends of the numerical error. The summary of our conclusions is presented in Sec.~IV.

\section{Numerical procedure}
\label{sec:numproc}

\subsection{Hamiltonian and target states}

In the QC-DMRG and QC-TTNS applications, the electron-electron correlation
is taken into account by an iterative procedure that minimizes the Rayleigh 
quotient corresponding to the Hamiltonian of the system: 
\begin{equation} 
\label{eqn:qcham}
H = \sum_{i j \sigma} T_{i j} c_{i \sigma}^{\dagger} c_{j \sigma} +
\sum_{i j k l \sigma \sigma'} V_{i j k l} c_{i \sigma}^{\dagger} c_{j \sigma'}^{\dagger} c_{k \sigma'} c_{l \sigma}.
\end{equation}
This Hamiltonian 
determines the exact states of the given molecule. In Eq.~(\ref{eqn:qcham}),
$c_{i \sigma}^{\dagger}$ and $c_{j \sigma}$ creates and annihilates 
an electron with spin $\sigma$, respectively.
$T_{i j}$ denotes the matrix elements of the one-particle
Hamiltonian, which is comprised of the kinetic energy and the external
electric field of the nuclei, and $V_{i j k l}$ stands for the matrix
elements of the electron repulsion operator, defined as
\begin{displaymath}
V_{i j k l} = \int d^3 x_1 d^3 x_2
\Phi_i^{*} (\vec{x}_1) \Phi_j^{*} (\vec{x}_2)
\frac{1}{\vec{x}_1-\vec{x}_2}
\Phi_k (\vec{x}_2) \Phi_l (\vec{x}_1) \, .
\end{displaymath}
The matrix elements $T_{ij}$ and $V_{ijkl}$ are
expressed in a molecular orbital (MO) basis obtained by
CASSCF optimizations.
The benchmark energies are computed with the same set of MO's.
(For more details, see section Basis states.)

In the present version of our TTNS method non-Abelian symmetries \cite{mcculloh,Toth2008,Zgid2008,Sharma2012,Wouters2013}
are not implemented yet, thus in order to specify the eigenstates we have fixed the 
number of electrons with up and down spins and shifted the triplet levels from the 
low lying spectrum by adding a term $\sum_{i,j} \delta \left(S_i^+S_j^- + S_i^-S_j^+\right)$ 
with $S_i^+=c_{i \sigma}^{\dagger} c_{j \sigma'}$ and $\delta=1$
to the Hamiltonian given in Eq.~(\ref{eqn:qcham}). 
Nevertheless, we have also checked the total spin of each states by calculating  
the expectation value of the 
$S^2 = \sum_{ij} S^-_i S^+_j + \sum_{ij} S^z_i S^z_j + \sum_i S^z_i\,$
operator which is equal to $S(S+1)$ in Hartree atomic units, i.e.,
zero for a singlet state and two for a triplet state.

In the MPS-based approaches, several eigenstates can be calculated within a single calculation.
Therefore, we have formed the reduced density matrix of the target state, $\rho$, from the 
reduced density matrices of the lowest $n$ eigenstates as $\rho = \sum_\gamma p_\gamma \rho_\gamma$ with 
$\gamma=1\ldots n$ and $p_\gamma=1/n$ at each bond length. 
In case of our QC-DMRG code the orbital spatial symmetry of the target 
state can also be fixed \cite{legeza2003a} in which case the 
first two lowest lying $^1\Sigma^+$ states 
can be calculated directly by using only $n=2$ eigenstates of the related subspace of
the Hamiltonian.
In the present version of our TTNS method, however, orbital spatial symmetry is not implemented yet. 
Therefore, we had to target the four lowest lying states with $n=4$ due to the fact that there 
are two additional eigenstates between the two lowest 
$^1\Sigma^+$  states. 
For more detailed descriptions of target states we refer to the original works and reviews 
\cite{white,dmrg_lif,schollwock2005}. 

\subsection{Basis states}
\label{sec:basis}

Atomic orbital (AO) basis was 
adopted from the work of Bauschlicher and  Langhoff\cite{fci_lif}
in order to match with previous DMRG computations 
\cite{dmrg_lif}.
The AO basis set of Ref. \cite{fci_lif} is suitable to describe 
the ionic and covalent LiF states as well. It
 consists of 9s and 4p functions contracted to
4s and 2p functions 
on the Li atom and 9s, 6p and 1d functions contracted to
4s, 3p and 1d
on the F atom. For more details of the AO basis set we
refer to the original publication\cite{fci_lif}. 

The two lowest $^1\Sigma^+$ states of LiF around the 
equilibrium bond length can be qualitatively described by the 
1$\sigma^2$2$\sigma^2$3$\sigma^2$4$\sigma^2$1$\pi^4$ and 
1$\sigma^2$2$\sigma^2$3$\sigma^2$4$\sigma^1$5$\sigma^1$1$\pi^4$ 
configurations\cite{fci_lif}. For this reason, the
 MO basis was obtained by CASSCF optimizations, with  two active electrons on
two active orbitals (4$\sigma$ and 5$\sigma$)  (CAS(2,2)).
MO's were optimized 
simultaneously for both $^1\Sigma^+$ states.
$T_{ij}$ and $V_{ijkl}$ matrix elements  of Eq.~(\ref{eqn:qcham}) 
are expressed in this MO basis.
CASSCF optimizations were
carried out with the GAMESS quantum chemistry package\cite{gamess}.

Orbitals 1$\sigma$, 2$\sigma$ and 3$\sigma$ were kept frozen
in all presented configurational interaction (CI), MPS(DMRG) and TTNS
computations. Six of the valence electrons were excited to all 
orbitals in the CI calculation, which we use as reference to compare
the TTNS results to. CI results were obtained
by utilizing the determinant-based full-configuration interaction
 (full-CI) program of Z. Rolik (Budapest), which is 
based on the CI algorithm of Olsen et al.\cite{ofci}.
C$_{2v}$ point group symmetry constraints were assigned during this study.

\subsection{The QC-TTNS method}
\label{sec:qcttns}

In this section, we present the brief overview of the most general QC-TTNS algorithm.
For the full description of the method we refer to the original work \cite{murg2010-tree}. 
In our implementation, we allow tensors to have orbital dependent coordination 
number, $z_i$,  in contrast to the implementation of 
Nakatani {\sl et. al.} \cite{chan2013-tree} which is an efficient extension of the 
DMRG method using fixed number of blocks.
Our main motivation is to develop an algorithm which reflects 
the entanglement structure of the molecule under study 
as much as possible (see Fig.~\ref{fig:I_free}).
The computational cost in one step of the algorithm scales as $D^{z_i+1}$ where $D$ 
is the dimension of the auxiliary space (in the DMRG community referred as block states)
while the the long-range correlation deviates from the
mean-field value only polynomially with distance. Therefore, there is a trade-off between
entanglement localization and increased order of the tensors. The construction of the
optimal tensor network is thus a far more complex task than it is in case of the MPS based approaches.

\begin{figure}[htb]
\centerline{
\scalebox{0.35}{\includegraphics{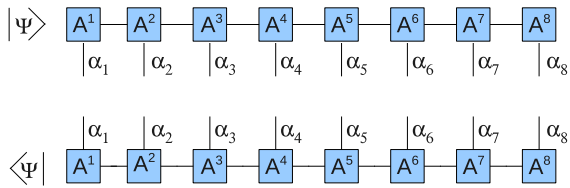}}
}
\caption{(Color online) Schematic plot of the matrix product state (MPS) algorithm. Each node is represented by a tensor of order 2
and the vertical line the physical index $\alpha$.}
\label{fig:mpsnetwork}
\end{figure}

In a full-CI treatment, a given eigenfunction of Eq.~(\ref{eqn:qcham}) can be written in a full 
tensor form as 
\begin{equation}
|\Psi\rangle = \sum_{\alpha_1,...,\alpha_N} U_{\alpha_1 \ldots \alpha_N} |\alpha_1,...,\alpha_N\rangle.
\end{equation}
where $U_{\alpha_1 \ldots \alpha_N}$ is a tensor with order $N$ and $|\alpha_i\rangle$ represents basis states
at molecular orbital $i$. In our study, a molecular orbital can be empty, singly occupied with up or down spins
or doubly occupied, thus the dimension of the local Hilbert space, $d$, is four. 
Since the number of independent parameters in $U$ scales exponentially with $N$ it is mandatory to
approximate such high dimensional tensor with a proper factorization in terms of lower dimensional tensors.  
In the MPS representation, $U$ describes a matrix network, i.e., it emerges from contractions of a set of matrices
$\{A^1,\ldots,A^N\}$, where
\begin{equation}
A^i \left [\alpha_i\right]_{m_{i-1},m_i} \, ,
\end{equation}
is a matrix at each vertex~$i$ of the network, with 
$2$ virtual indices $m_{i-1}, m_i$ of dimension~$D$ and one physical
index~$\alpha_i$ of dimension~$d$, thus  
\begin{equation}
|\Psi\rangle = \sum_{\alpha_1,...,\alpha_N}^d A_{\alpha_1}^1 A_{\alpha_2}^2 \ldots A_{\alpha_N}^N |\alpha_1\rangle|\alpha_2\rangle...|\alpha_N\rangle.
\end{equation}
The schematic plot of the matrix product state (MPS) network is shown in Fig.~\ref{fig:mpsnetwork}.

\begin{figure}[htb]
\centerline{
\scalebox{0.20}{\includegraphics{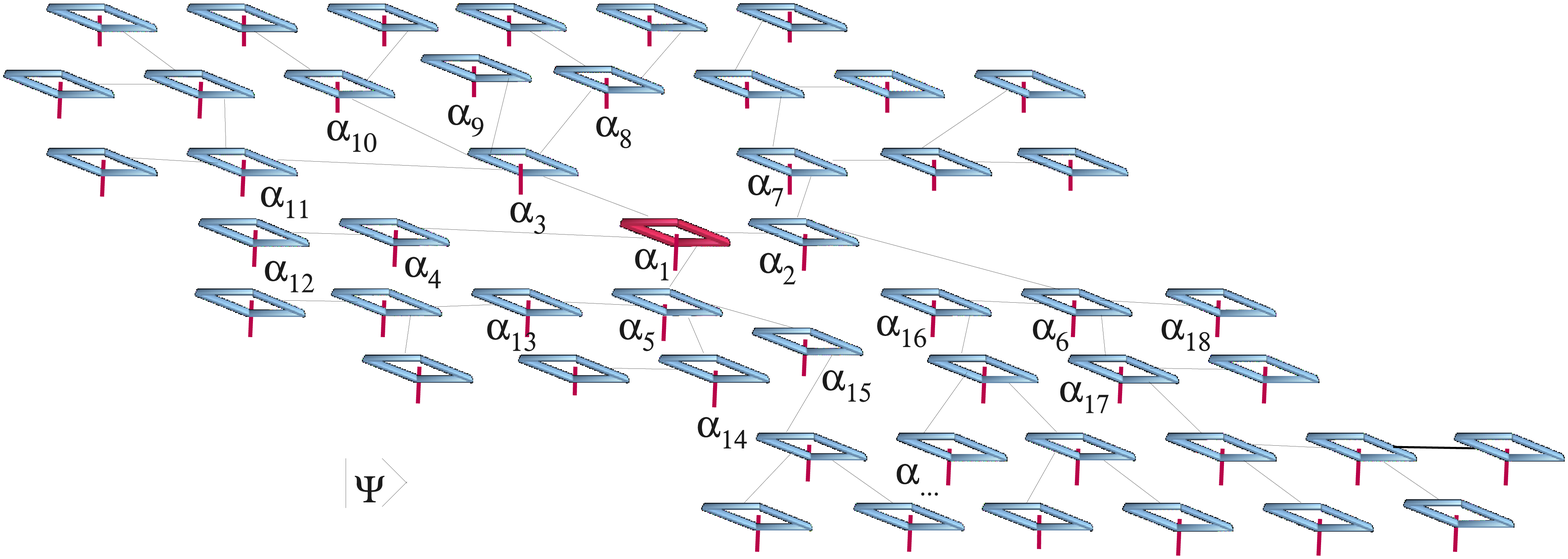}}
}
\vskip 0.2cm
\centerline{
\scalebox{0.20}{\includegraphics{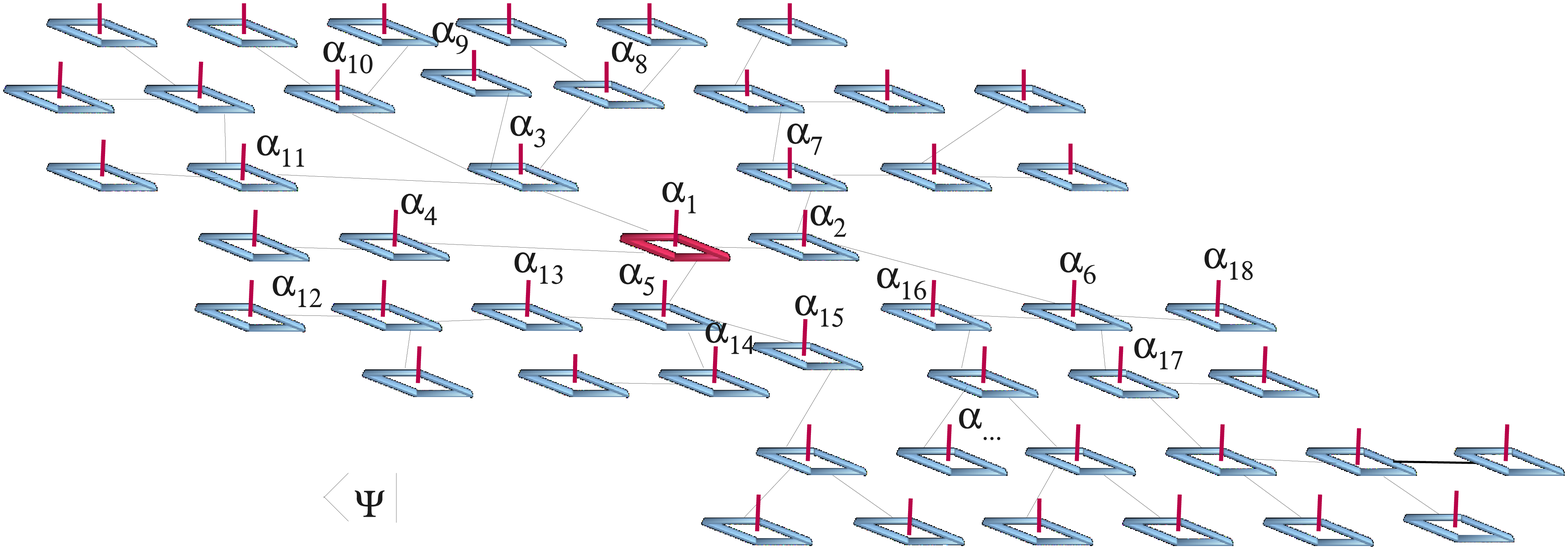}}
}
\caption{(Color online) Schematic plot of a higher dimensional network, for example, the tree tensor network state (TTNS) algorithm.
Each node is represented by a tensor of  order $z_i$, where
$z_i$ is an orbital dependent coordination number. The network supposed
to reflect the entanglement structure of the molecule as much as possible.
The vertical red lines denote physical indices $\alpha_i$, $i\in \{1,N\}$.
Entanglement is transferred via the virtual bonds that connect the orbitals shown by black lines.
The central node is indicated by red contour.}
\label{fig:ttns}
\end{figure}

\begin{figure}[htb]
 \centerline{
\scalebox{0.5}{\includegraphics{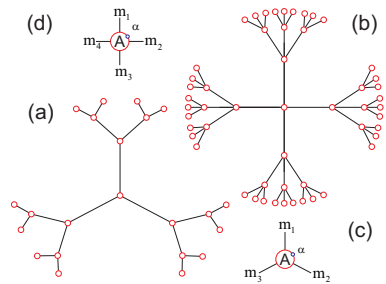}}
 }
\caption{(Color online) Top view of the tree tensor network (TTNS) 
algorithm with fixed coordination number, (a) $z_i=3$, and (b) $z_i=4$.
The structure of the 
tensors is shown in (c) and (d). The bonds indicate the virtual indices 
$m_1,\ldots,m_z$ and the circle the physical index~$\alpha$.
} 
\label{fig:treenetwork}
\end{figure}

A natural extension of the MPS approach is to use higher order tensors.
In this work, we form a tree tensor
network in which all sites in the tree represent physical orbitals 
and in which entanglement is transferred via the virtual bonds that
connect the sites as shown in Fig.~\ref{fig:ttns}.
Our motivation is to treat models in which orbitals have varying
degrees of entanglement; positions closer to the center of the tree
should be better suited to represent more entangled sites.
An additional motivation is to take advantage of the property of the
tree tensor network ansatz that the long-range correlations differ 
from the mean-field value polynomially with distance rather than
exponentially with distance as for MPS. In our algorithmic approach to 
optimize the tree tensor network, we use tools similar to those used in
Refs.~[\onlinecite{shi2006}],
~[\onlinecite{Tagliacozzo2007}],
~[\onlinecite{Corboz2009}], and ~[\onlinecite{Corboz2010}],
and optimize the network site-by-site as in the DMRG.
Therefore, 
$U_{\alpha_1 \ldots \alpha_N}$ can describe a tree tensor
network, i.e., they emerge from contractions of a set of tensors
$\{A_1,\ldots,A_N\}$, where
\begin{equation}
A^i\left[ \alpha_i \right]_{m_1 \ldots m_z} \, ,
\end{equation}
is a tensor with $z+1$ indices, at each vertex~$i$ of the network 
according to Fig.~\ref{fig:treenetwork}.
Each tensor has
$z$ virtual indices $m_1 \ldots m_z$ of dimension~$D$ and one physical
index~$\alpha_i$ of dimension~$d$, with $z$ being the coordination number of that site.
The coefficients $U_{\alpha_1 \ldots \alpha_N}$ are obtained by contracting the
virtual indices of the tensors according to the scheme of a tree
tensor network (see Fig.~\ref{fig:treenetwork}). The structure of the
network can be arbitrary and the coordination number can vary from
site to site. The only condition is that the network is bipartite,
i.e., by cutting one bond, the network separates into two disjoint
parts.
Therefore, the TTNS network does not contain any loop (see Fig.~\ref{fig:treenetwork})
which allows an exact mathematical treatment \cite{hackbusch,schneider-rev}.
For $z=2$, the one-dimensional MPS-ansatz used in DMRG is recovered.

Since entanglement is transferred via the virtual bonds
that connect the sites, it is preferable to put strongly
correlated sites close together, i.e. to minimize the number of bonds between them.
For $z>2$ the number of virtual bonds required to connect two arbitrary orbitals scales logarithmically
with the number of orbitals~$\alpha$, whereas the scaling is linear in~$N$ for $z=2$.
This can be seen by considering a Cayley-tree of depth~$\Delta$, as shown in Fig.~\ref{fig:treenetwork}.
The number of sites in the tree is
\begin{displaymath}
N = 1+z \sum_{j=1}^{\Delta} (z-1)^{j-1} = \frac{z(z-1)^{\Delta}-2}{z-2}
\end{displaymath}
and thus, the maximal distance between two orbitals, $2 \Delta$, scales logarithmically with~$N$ for $z>2$.
Because of this logarithmic scaling, the expectation value of a long-range correlations differs
from the mean-field result by a quantity that scales polynomially with distance.
This contrasts the MPS ansatz ($z=2$) that shows an exponential decay of the difference with distance.~\cite{murg2010-tree}

The TTNS algorithm consists in the variational optimization of the tensors $A_i$
in such a way that the energy is minimized (with the constraint that the norm of the state remains constant).
This is equivalent to 
optimizing the functional
\begin{displaymath}
F = \bra{\Psi} H \ket{\Psi} - E \left( \scal{\Psi}{\Psi} - 1 \right),
\end{displaymath}
where $\Psi=\Psi(A_1,\ldots, A_N)$.
This functional is non-convex with respect to all parameters
$\{A_1,\ldots,A_N\}$. However, fixing all tensors $A_k$ except $A_i$,
due to the tensor network structure of
the ansatz, it is quadratic in the parameters $A_i$ associated with
one lattice site~$i$. Because of this, the optimal parameters $A_i$
can simply be found by solving a generalized eigenvalue problem
$\mathcal{H}_i \vec{A}_i = E \mathcal{N}_i \vec{A}_i$. For a bipartite
network, it is always possible to assume a gauge condition so that
$\mathcal{N}_i = \mathbb{1}$, and thus reduce the generalized
eigenvalue problem to an ordinary one.~\cite{murg2010-tree}
The concept of the algorithm is to do
this one-site optimization site-by-site until convergence is reached.
The challenge that remains is to calculate the effective Hamiltonian
$\mathcal{H}_i$ of the eigenvalue problem. In principle, this is done
by contracting all indices in the expression for the expectation value
$\bra{\Psi} H \ket{\Psi}$ except those that connect
to $A_i$. By
interpreting the tensor $A_i$ as a $q D^z$-dimensional vector
$\vec{A}_i$,
this expression can be written as
$\bra{\Psi} H \ket{\Psi} = \vec{A}_i^{\dagger} \mathcal{H}_i \vec{A}_i$.
Since
$\scal{\Psi}{\Psi} = \vec{A}_i^{\dagger} \mathcal{N}_i \vec{A}_i$
and $\mathcal{N}_i = \mathbb{1}$,
the functional~$F$ attains its minimum when
\begin{displaymath}
\mathcal{H}_i \vec{A}_i = E \vec{A}_i.
\end{displaymath}
Due to the bipartite structure of the tensor network, the calculation
of $\mathcal{H}_i$ can be performed efficiently, i.e., on
a time that scales polynomially with~$N$ and~$D$.

\begin{figure}[t]
    \begin{center}
        \includegraphics[width=0.44\textwidth]{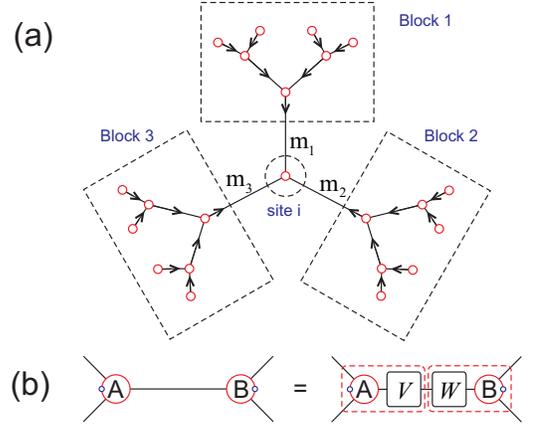}
    \end{center}
    \caption{
       (Color online) (a) Separation of the state into~$z$ blocks plus
      the site under optimization, as described by
      Eq.~(\ref{eqn:blockwavefct}). (b) Natural freedom in the tensor
      network: insertion of a matrices $V$ and $W$
      fulfilling $V W = \mathbb{1}$
      at one bond
      leaves the state invariant. The contraction of $A$ with $V$
      forms the new tensor~$A'$ on the left hand side; the contraction
      of $B$ with $W$ forms the new tensor~$B'$ at the right hand
      side.
        }
    \label{fig:orthonormalization}
\end{figure}
This TTNS algorithm is similar to a
DMRG calculation with~$z$ blocks instead of two,
where a block consists of all of the sites within one of the branches
emerging from site~$i$ (see Fig.~\ref{fig:orthonormalization}(a)).
The wave function is then formed as
\begin{equation} \label{eqn:blockwavefct}
\ket{\Psi} =
\sum_{m_1,\ldots,m_z=1}^D
\ket{\varphi_{m_1 \ldots m_z}} \otimes
\ket{\phi_{m_1}^1} \otimes \cdots \otimes \ket{\phi_{m_z}^z}
\, ,
\end{equation}
where
$\ket{\phi_{m}^{\gamma}}$ ($m=1,\ldots,D$) is
the basis in environment block~$\gamma$ ($\gamma=1,\ldots,z$) and
$\ket{\varphi_{\alpha_1 \ldots \alpha_z}}$ is the state of
site~$i$.
Matrix $\mathcal{N}_i$ is equal to the identity if
the basis $\ket{\phi_{m}^j}$ in each environment block
is orthonormal.
This can always be achieved, because of a gauge degree of
freedom in a TTNS\cite{murg2010-tree,Haegeman2013,Lubich}: it is possible to insert at any bond
a resolution of the identity $\mathbb{1} = V W$,
and contract the matrices $V$ and $W$ with the adjacent
tensors $A_i$ (see Fig.~\ref{fig:orthonormalization}b).
This does not change the state, but changes
the tensors $A_i$ and the basis states
$\ket{\phi_{m}^j}$ of the environment blocks.
It can be shown that it is always possible to find
gauge transformations that orthonormalize the
basis states of the environment blocks.~\cite{murg2010-tree}

\begin{figure}[t]
    \begin{center}
        \includegraphics[width=0.44\textwidth]{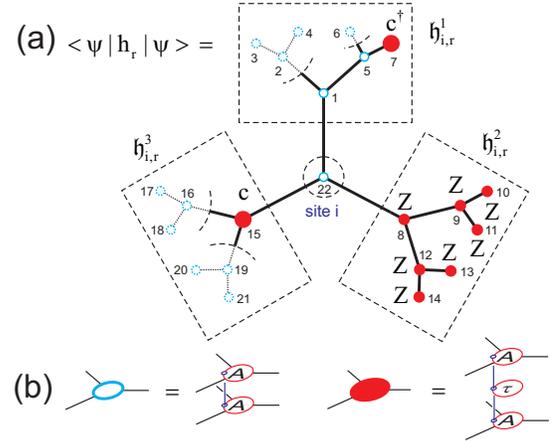}
    \end{center}
    \caption{
       (Color online) Formation of the effective Hamiltonian $\mathfrak{h}_{i,r} = \mathfrak{h}_{i,r}^1 \otimes \mathfrak{h}_{i,r}^2 \otimes \mathfrak{h}_{i,r}^3$ with respect to the fermionic interaction $h_r = c_7^{\dagger} c_{15}$.
Between sites $7$ and $15$ a chain of $Z$-matrices appears due to the
Jordan-Wigner transformation.
The sites on which the interaction has support are marked in red.
Each open (filled) circle in the tensor network corresponds to the 
contraction of the layered structure of tensors shown in (b).
        }
    \label{fig:heff}
\end{figure}

Thus, by assuring that the gauge condition is always satisfied in the
course of the algorithm, the only term that must
be calculated is
the effective Hamiltonian~$\mathcal{H}_i$. This term is obtained
by contracting all tensors
except $A_i$ in the expectation value~$\bra{\Psi} H \ket{\Psi}$.
Hamiltonian~$H$ is a sum of $O(N^4)$
two-point and four-point fermionic interaction terms
$c_{i \sigma}^{\dagger} c_{j \sigma}$ and
$c_{i \sigma}^{\dagger} c_{j \sigma'}^{\dagger} c_{k \sigma'} c_{l \sigma}$
($\sigma,\sigma' \in \{\uparrow,\downarrow\}$),
as defined in eq.~(\ref{eqn:qcham}).
The fermionic nature of the terms can be handled by mapping them to
non-local bosonic operators via Jordan-Wigner transformations.
Writing $H = \sum_r h_r$
with $h_r$ denoting the Jordan-Wigner transformed two and four-point interaction terms,
the effective Hamiltonian~$\mathcal{H}_i$ transforms into a sum
$\mathcal{H}_i = \sum_r \mathfrak{h}_{i,r}$, where
\begin{displaymath}
\bra{\Psi} h_r \ket{\Psi} = \vec{A}_i^{\dagger} \mathfrak{h}_{i,r} \vec{A}_i \, .
\end{displaymath}
Due to the structure~(\ref{eqn:blockwavefct}) of the TTNS, each
effective Hamiltonian~$\mathfrak{h}_{i,r}$ factorizes into a tensor product of~$z$ matrices
\begin{displaymath}
\mathfrak{h}_{i,r} = \mathfrak{h}_{i,r}^1 \otimes \cdots \otimes
\mathfrak{h}_{i,r}^z \, ,
\end{displaymath}
where each matrix $\mathfrak{h}_{i,r}^{\gamma}$ corresponds
to the matrix elements of $h_i$ with respect to the basis in environment block~$\gamma$:
\begin{displaymath}
\left[ \mathfrak{h}_{i,r}^{\gamma} \right]_{m n} =
\bra{\phi_m^{\gamma}} h_r \ket{\phi_n^{\gamma}} \, .
\end{displaymath}
Graphically, the evaluation of $\bra{\phi_m^{\gamma}} h_r \ket{\phi_n^{\gamma}}$
corresponds to the contraction of a three-layered
tensor network according to the structure of the branch in block~$\gamma$,
as depicted in Fig.~\ref{fig:heff}. This network can be contracted
efficiently by starting from the leaves and working in the inward direction.

With TTNS we can easily enforce the $U(1)$ symmetry that is
fulfilled by Hamiltonian~(\ref{eqn:qcham}),
i.e. the conservation of the number of particles.
For this, the tree graph has to be made
directed (see Fig.~\ref{fig:orthonormalization}a), such that all sites
(except site~$i$ that is optimized)
have~$z-1$ incoming and one outgoing bond.
Thus, each virtual index of
a tensor~$A_i$ is equipped with the additional information of whether
it is ``incoming'' or ``outgoing''.
Each virtual index connecting to block~$\gamma$ ($\gamma=1,\ldots,z$)
is split into an index tuple $(m_{\gamma},n_{\gamma}^{\uparrow},n_{\gamma}^{\downarrow})$.
Assuming that the index connecting to block~$\gamma=1$ is
the outgoing index,
we require that
$n_1^{\uparrow} = n_2^{\uparrow} + \ldots + n_z^{\uparrow} + n^{\uparrow}(\alpha)$ and
$n_1^{\downarrow} = n_2^{\downarrow} + \ldots + n_z^{\downarrow} + n^{\downarrow}(\alpha)$.
$n^{\uparrow}(\alpha)$ ($n^{\downarrow}(\alpha)$) denotes the
number of electrons with spin up (down) corresponding to the physical index~$\alpha$.
Thus, for $\gamma=2,\ldots,z$,
$n_{\gamma}^{\uparrow}$ ($n_{\gamma}^{\downarrow}$) counts the number
of up-electrons (down-electrons) within branch~$\gamma$.
The index $n_1^{\uparrow}$ ($n_1^{\downarrow}$), on the other hand,
is equal to the number
of electrons with spin up (down) in all the branches
plus the number of electrons at site~$i$.

\begin{figure}[t]
    \begin{center}
        \includegraphics[width=0.44\textwidth]{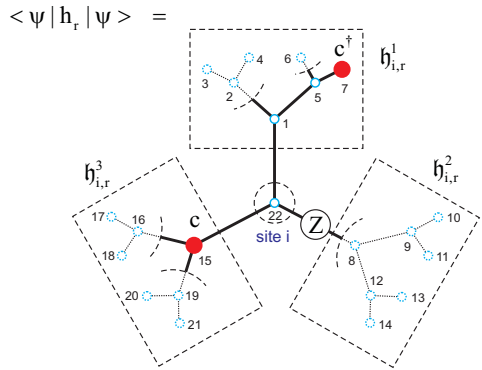}
    \end{center}
    \caption{
       (Color online) Formation of the effective Hamiltonian $\mathfrak{h}_{i,r} =
        \mathfrak{h}_{i,r}^1 \otimes \mathfrak{h}_{i,r}^2 \otimes
        \mathfrak{h}_{i,r}^3$ with respect to the fermionic
        interaction $h_r = c_7^{\dagger} c_{15}$ with particle number conservation taken into account. The sites on which the interaction has support are marked in red. All branches marked by dotted lines and circles yield the identity when contracted. The parity operator~$\tilde{Z}$ is contracted to the virtual bond.
        }
    \label{fig:hefffermionswithparity}
\end{figure}
Besides the ability of targeting a state
with a specific total number of up- and down-electrons $N_{\uparrow}$ and $N_{\downarrow}$
and giving a performance boost to the algorithm
(the virtual dimension effectively increases from~$D$
to $D (N_{\uparrow}+1) (N_{\downarrow}+1)$),
the inclusion of the particle-number conservation
also simplifies the treatment of the fermionic nature
of the electrons.
The main idea is depicted in
Fig.~\ref{fig:hefffermionswithparity} for the
interaction~$c_7^{\dagger} c_{15}$: with an appropriately chosen
numbering of the fermions, each subbranch that has no fermionic
support either has only
identities acting on the sites or only
matrices~$Z$ stemming from the Jordan-Wigner transformation
($Z_{\alpha \beta} = 
\delta_{\alpha \beta} (-1)^{n^{\uparrow}(\alpha)+n^{\downarrow}(\alpha)}$).
The subbranches including only identities simplify to
the identity because we work in a gauge
in which the basis in each environment block
is orthonormal.
As shown in~[\myonlinecite{murg2010-tree}], the
$Z$ operators can be ``moved'' to the virtual bonds
and, since $Z^2=\mathbb{1}$, all of them except one cancel (see Figs.~\ref{fig:heff}
and~\ref{fig:hefffermionswithparity}).
What remains is a subbranch
that includes
only identities, which reduces to
the identity because of the
orthonormalization of the state. Thus, for a fermionic two-site
interaction, it is sufficient to take into account the path connecting
the two sites.
In this way,
the treatment of long-range fermionic
interactions is feasible with the same numerical effort as the
treatment of long-range spin interactions.

The numerical effort of the algorithm has two major contributions.
On the one hand, the bond-dimension $D$ is crucial: the numerical effort
for calculating one term of the effective Hamiltonian by tensor contraction
scales as $D^{z+1}$.
On the other hand, this calculation has to be performed for each
term in the Hamiltonian, such that naively a scaling $N^4 D^{z+1}$
is expected.
Fortunately, the same summation tricks as described in~[\myonlinecite{white-qc}]
can be applied, such that the scaling reduces to $N^2 D^{z+1}$.
Since $O(N)$ iteration steps are required for convergence,
the overall time of the algorithm will scale as $N^3 D^{z+1}$.

\subsection{Network optimization by entanglement localization}
\label{sec:optim}

\begin{figure}[htb]
        \centerline{
        \includegraphics[scale=0.37]{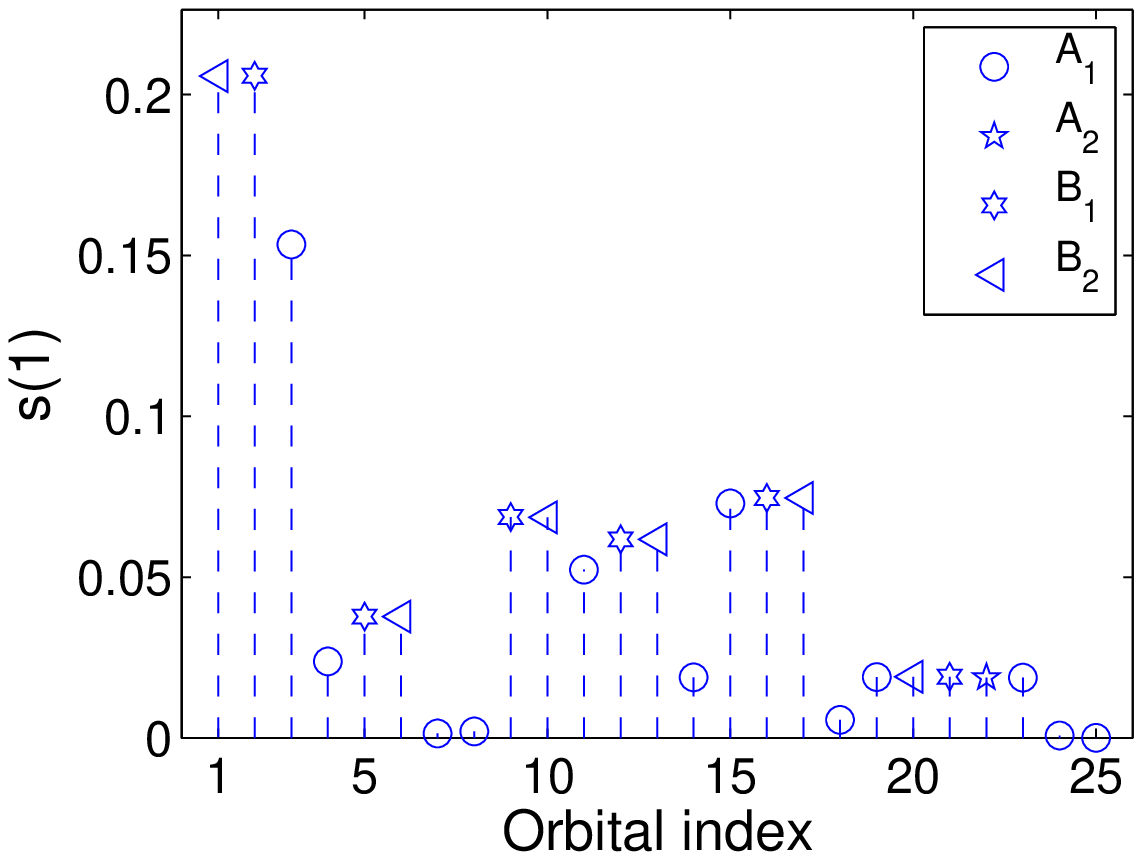}
        \hskip -0.3cm
        \includegraphics[scale=0.37]{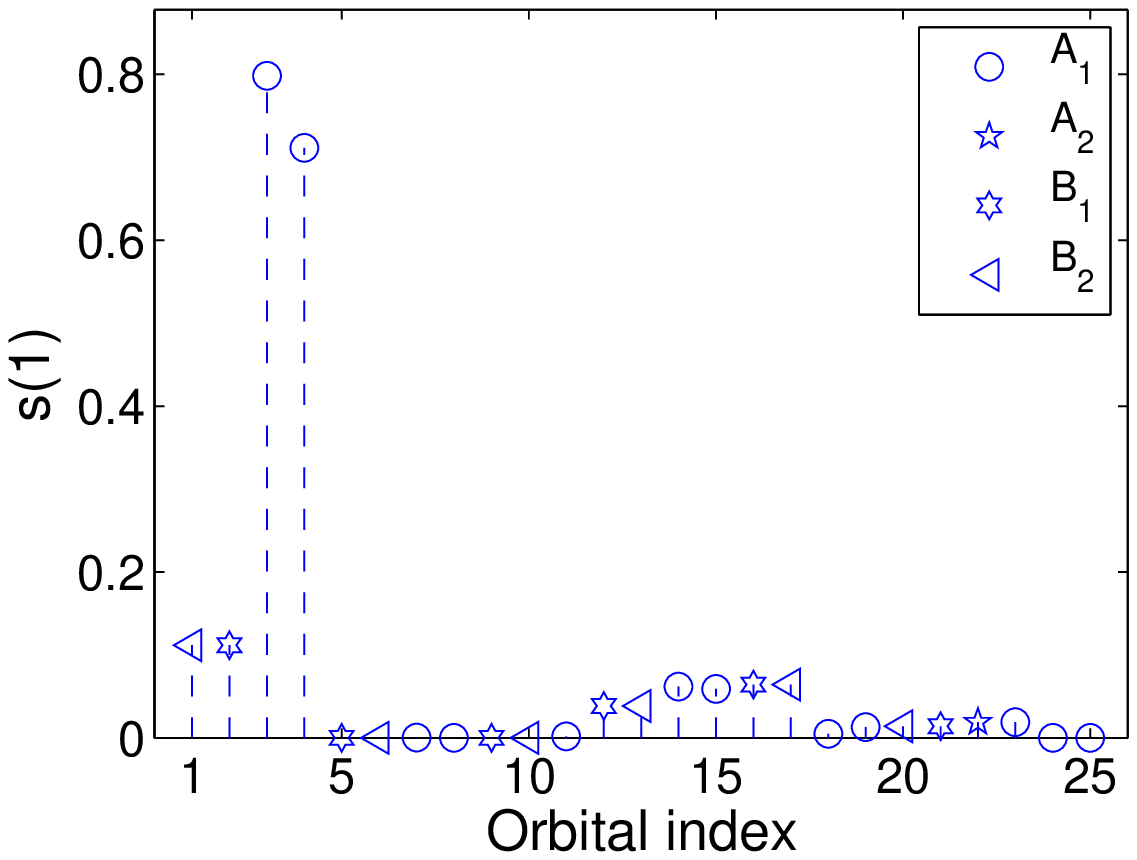}
        }
    \caption{(Color online) One orbital entropy profile for the LiF molecule at bond length (a) $r=3.05$ and at (b) $r=13.7$.
    Symbols label the irreducible representations of the molecular orbitals in the C$_{2v}$ point group.
    }
    \label{fig:s_i}
\end{figure}

\begin{figure}[htb]
        \centerline{
        \includegraphics[scale=0.26]{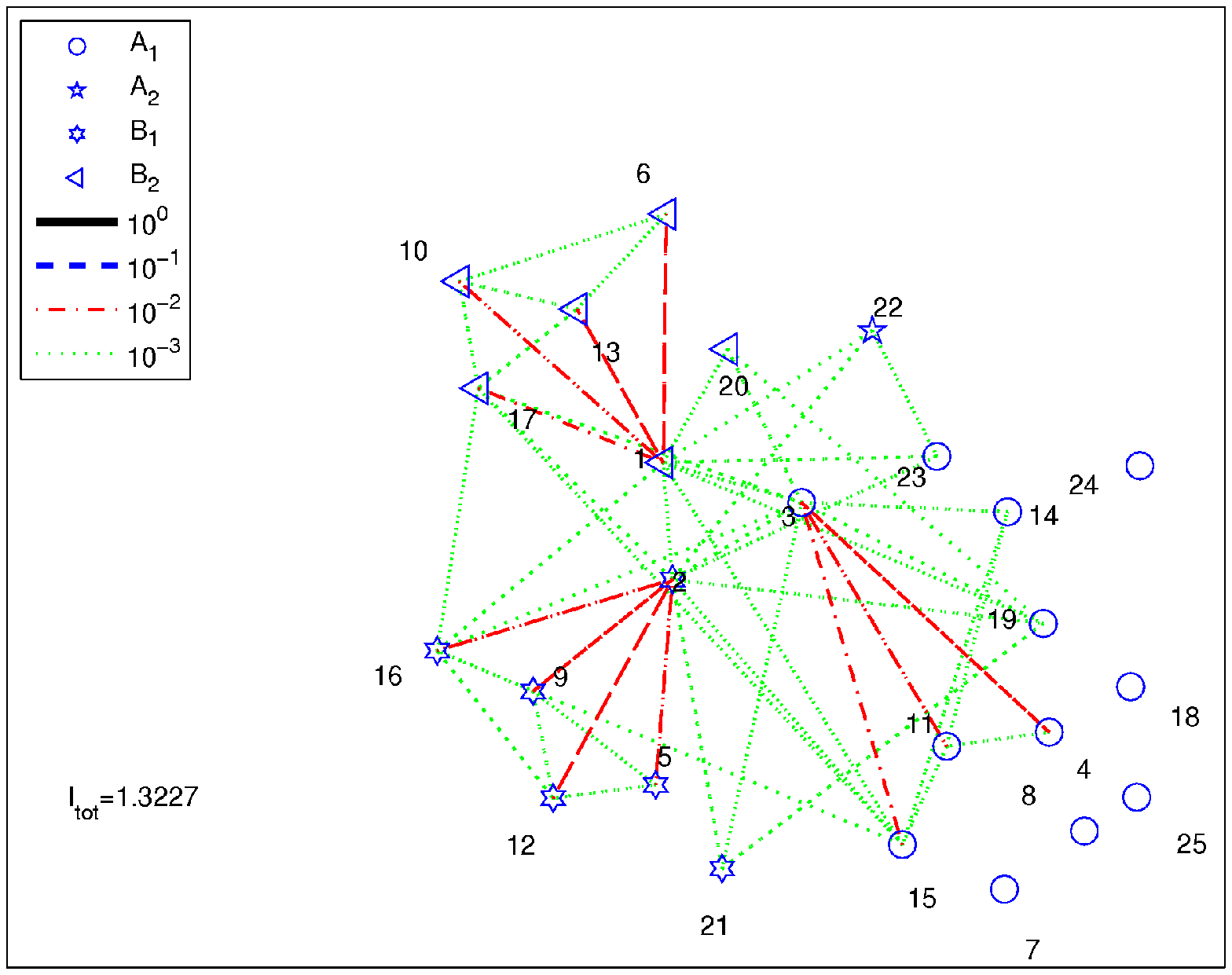}
        \includegraphics[scale=0.26]{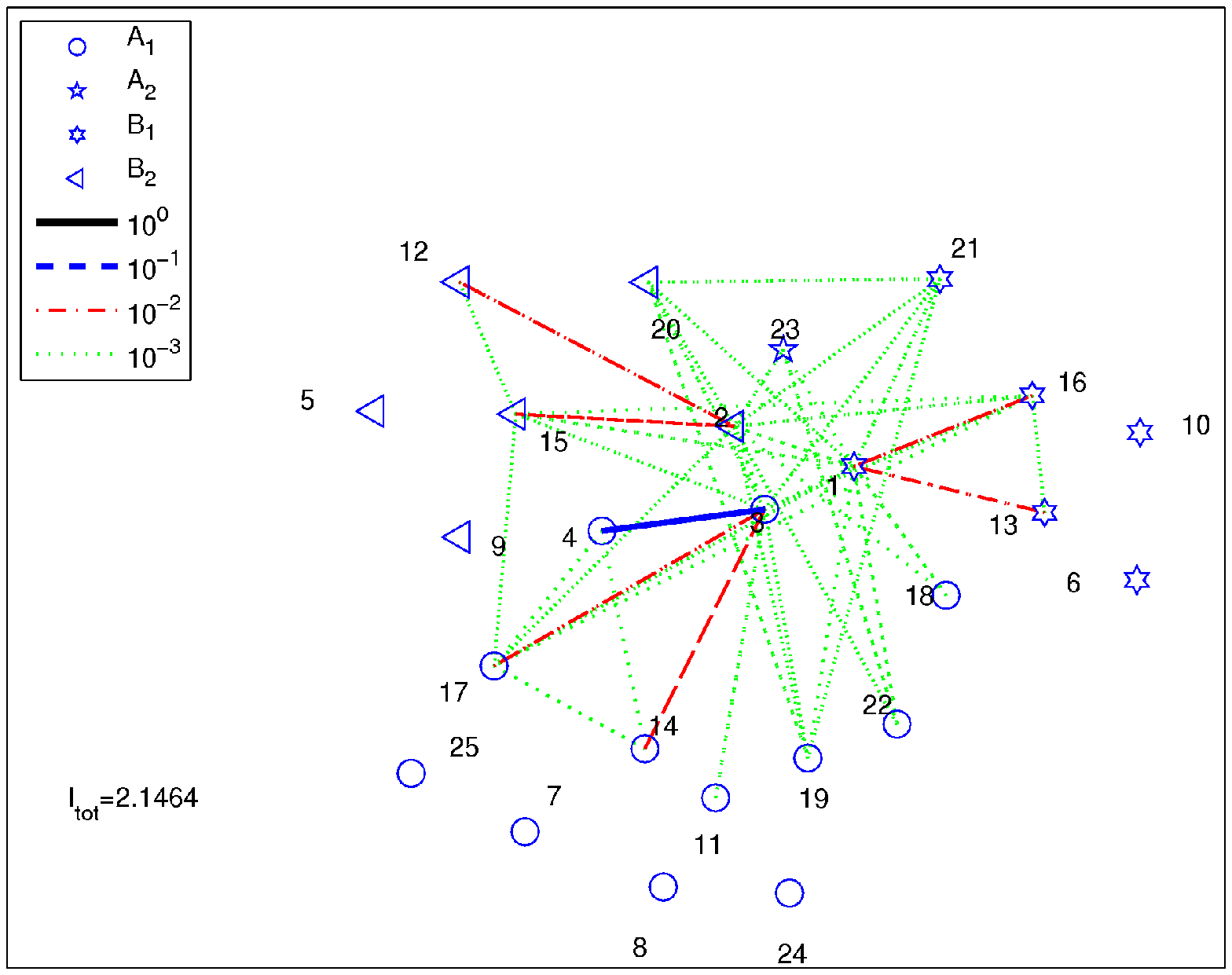}
        }
    \caption{(Color online) Mutual information represented as a two-dimensional 
weighted graph for the LiF molecule at bond length  
    (a) $r=3.05$ and at (b) $r=13.7$. Colors indicate different strengths of $I_{ij}$ and 
    the symbols label the irreducible representations of the molecular orbitals in the C$_{2v}$ point group.
    }
    \label{fig:I_free}
\end{figure}

The amount of contribution to the total correlation energy of an orbital can be detected qualitatively 
by the single-orbital entropy, $s(1)_i = -{\rm Tr} \rho_i \ln \rho_i$ where $\rho_i$ is the 
reduced density matrix at orbital $i$.
The two-orbital entropy is constructed similarly using the reduced density matrix, $\rho_{ij}$ of a subsystem 
built from orbitals $i$ and $j$ and the 
mutual information $I_{ij}=s(2)_{ij}-s(1)_i-s(1)_j$ 
describes how orbitals are entangled with each other as they are embedded in the whole system. 
For more detailed derivations we refer to the original papers \cite{legeza03c,rissler06,barcza2010a,boguslawski2013a}.
Therefore, these quantities provide chemical information about the system, especially about bond formation and nature of static 
and dynamic correlation \cite{boguslawski2012b,boguslawski2013a,boguslawski2013b,yanai2013}. 
As an example, $s(1)_i$ and $I_{ij}$ are shown in Figs.~\ref{fig:s_i} and \ref{fig:I_free}, respectively, 
for the equilibrium bond length $r=3.05$ and at large separation $r=13.7$. 
The total quantum information encoded in the wavefunction is given by the sum of the orbital entropy, i.e.,
$I_{\rm tot}=\sum_i s(1)_i$
which is twice as large for the stretched geometry as compared to the equilibrium case. 

It is clear form Fig.~\ref{fig:I_free} that some orbitals are strongly entangled 
with several other orbitals while some orbitals are entangled with only a few others 
and some are almost disentangled from the system. Therefore, the obvious choice
is to allow the coordination number, $z_i$, to vary from orbital to orbital. In the following analysis, however, 
we restrict ourself to a fixed $z_i=3$ case in order to allow a more direct analysis when data 
are compared to the $z_i=2$ MPS case. 

\begin{figure}[htb]
        \centerline{
        \includegraphics[scale=0.27]{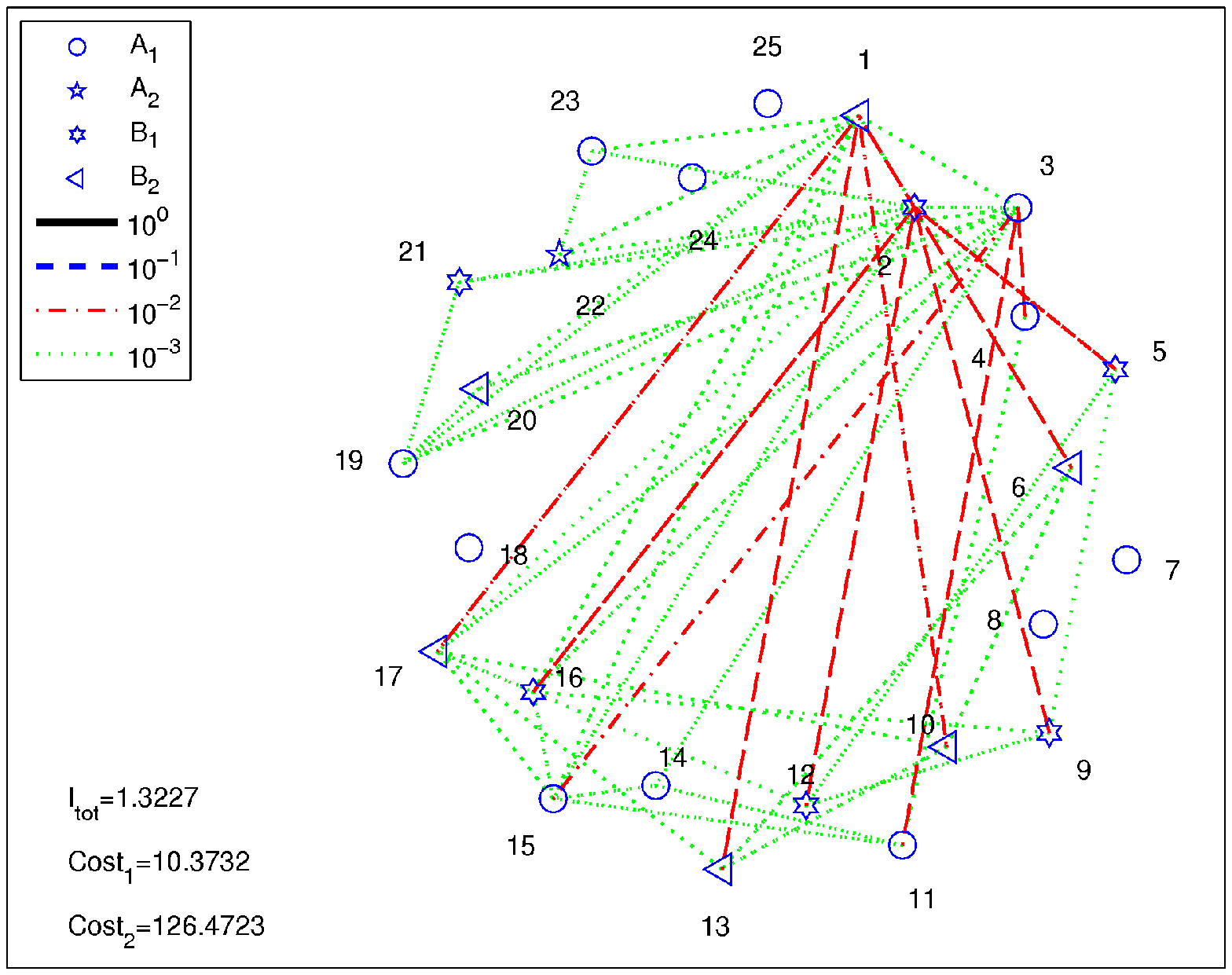}
        \includegraphics[scale=0.27]{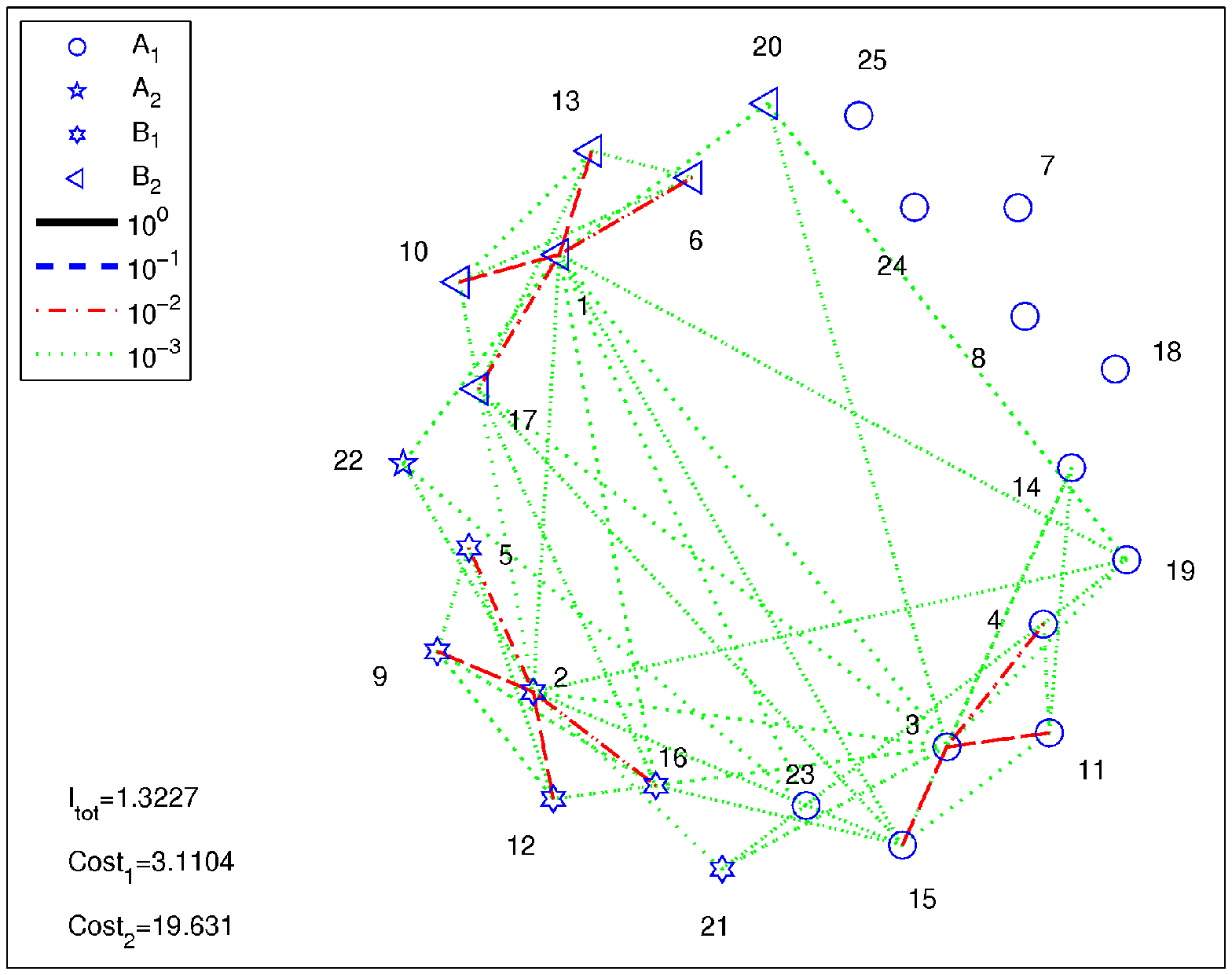}}
        \centerline{
        \includegraphics[scale=0.37]{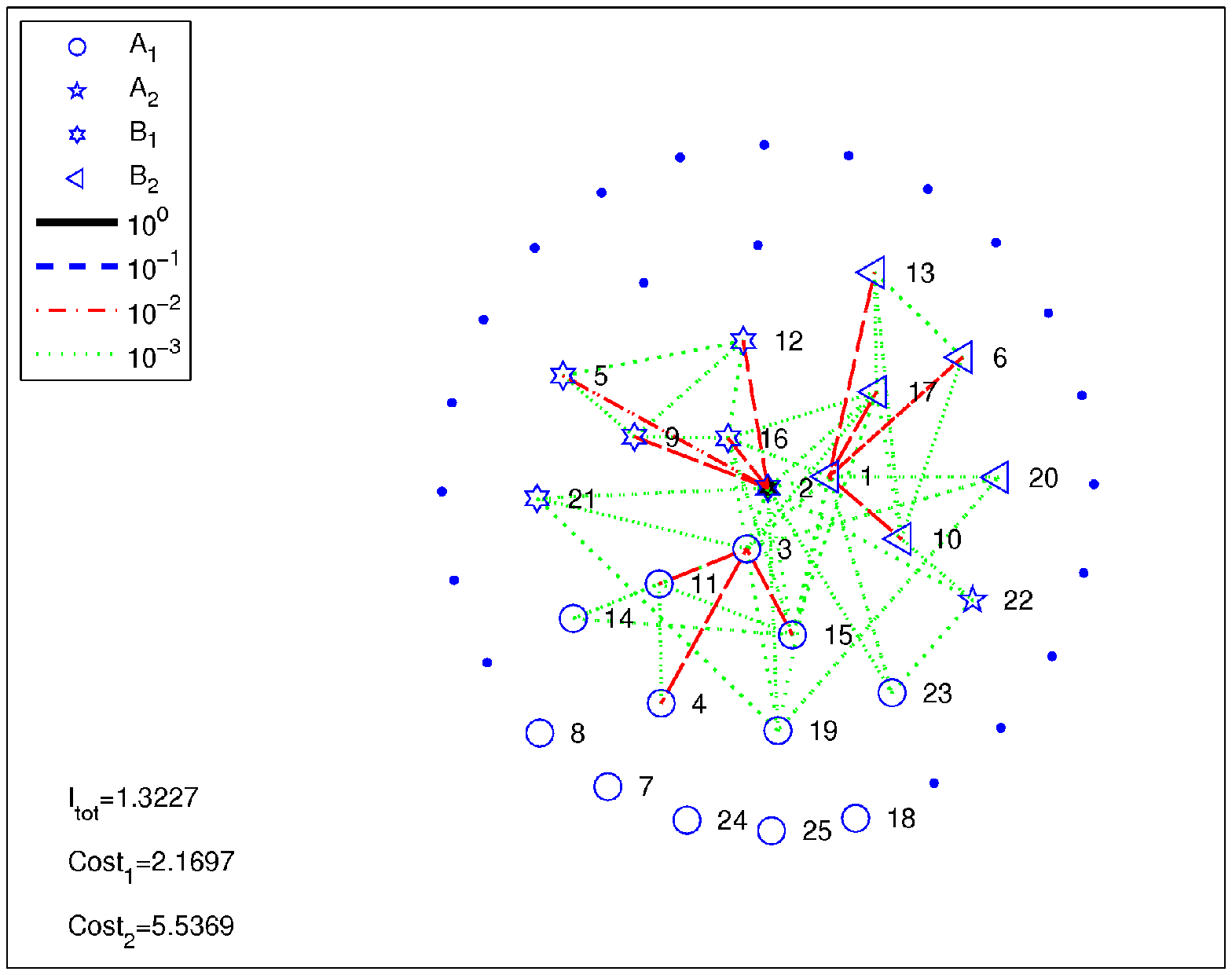}
        }
    \caption{(Color online) Optimization of tensor topology by minimizing the entanglement length in the system at
the equilibrium bond length $r=3.05$. 
(a) and (b) are for the one dimensional MPS like topology for the original ordering and for the optimized ordering, respectively. 
(c) Shows the optimized topology on the tree (small dots indicate not used grid points of the tree). The total quantum information 
$I_{\rm tot}$ does not change but the entanglement length calculated with $\eta=1$
and $2$ indicated by ${\rm Cost}_1$ and ${\rm Cost}_2$ drops significantly.
    }
    \label{fig:I_opt}
\end{figure}

Since both DMRG and TTNS rely on the systematic application of the Schmidt-decomposition the required 
computational resources to reach a given error margin is determined by the amount of entanglement 
in the system\cite{legeza2003a}.
This can be manipulated by changing the basis functions \cite{legeza2005-basis,murg2010-tree} or by changing the tensor topology.
For the latter case, the entanglement length, 
\begin{equation}
{\rm Cost}_\eta = \sum_{ij} I_{ij}\times d_{ij}^{\eta}, 
\label{eq:cost}
\end{equation}
should be minimized
in order to localize the entanglement in the system, where $d_{ij}$ is the distance function between orbital $i$ and $j$
depending on the tensor topology and $\eta$ is some exponent that we set to $1$ or $2$. For the one-dimensional 
case, i.e., for DMRG and MPS $d_{ij}=|i-j|$. For the tree topology 
$d_{ij}$ can be computed as the distance from the center to $i$, plus the distance from the center to $j$,
minus twice the distance from the center to their lowest common ancestor.
The lowest common ancestor can be obtained within a linear preprocessing time $O(N)$ and a constant query time
using the Berkman's algorithm.~\cite{berkman93}
As an example, the optimized tensor topologies
at the equilibrium bond length $r=3.05$ are shown in Figs.~\ref{fig:I_opt} for the one-dimensional topology and for the tree topology.  

In practice, first we performed a quick and fast DMRG full sweep with a fixed small number of block states 
($M\simeq16,\ldots,64$) using the ordering of orbitals for which the $T_{ij}$ and $V_{ijkl}$ integral files 
were generated by the GAMESS program in order to determine the one and two-orbital entropy profiles qualitatively. 
For all subsequent calculations we rendered orbitals with descending orbital entropy values to
form the Complete Active Space CAS-vector that was used during the 
configuration interaction based dynamic extended active space (CI-DEAS) procedure\cite{legeza03c}.
We also determined the mutual information, $I_{ij}$,
and the orbitals were reordered by minimizing the entanglement length given by Eq.~(\ref{eq:cost}). In the one-dimensional case
either the graph Laplacian can be used or by other heuristic methods to reduce the spectral envelope of $I_{ij}$, i.e., to make it
as diagonally dominant as possible\cite{fiedler1,fiedler2}. In case of the tree network, the optimization is less straightforward, 
but as a rule of thumb 
we placed orbitals with largest entropy values close to the center of the network
while keeping orbitals with large $I_{ij}$ values 
close together (see Fig.~\ref{fig:I_opt}(c)). 
In the subsequent step, accurate DMRG (with optimized CAS vector) 
or MPS/TTNS calculations were performed.

\section{Numerical results}
\label{sec:numres}

\subsection{Ground state and excited states}

In this work we have used two codes. Our QC-DMRG program has been developed for a long time
and it includes advanced features like the dynamic block state selection (DBSS) approach 
\cite{legeza2003a,legeza04a}, the CI-DEAS procedure\cite{legeza03c}, and the
treatment of orbital spatial symmetries which are not implemented yet in the TTNS code. Therefore,
the entropy functions were calculated by the QC-DMRG code to provide initial data quickly for network optimizations.
The rest of the analysis for a fair treatment is, however, based on the TTNS code alone using 
fixed $z_i=2$ (MPS) and $z_i=3$ (TTNS) coordination number while keeping all other 
parameters of the algorithm the same.  

\begin{figure}[htb]
        \centerline{
        \includegraphics[scale=0.5]{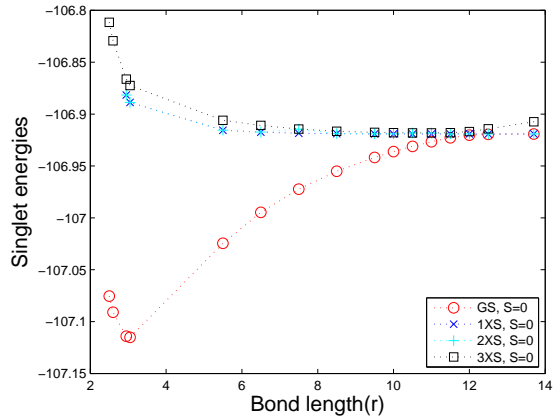}
        }
    \caption{(Color online) The energy of the four lowest lying states as a function bond length.} 
    \label{fig:pes_dmrg}
\end{figure}

The potential energy curve (PES) can be calculated for an a priory set error margin 
using the DBSS procedure \cite{dmrg_lif}. Therefore, we have easily reproduced the full-CI energies
up to $10^{-8}$ a.u. in  absolute error. 
In the following, however, we have used a small
fixed number of block states, i.e., fixed bond dimension, in order to demonstrate the
benefits underlying the TTNS geometry. The four lowest lying eigenstates ($\gamma=1,\ldots,4$) 
are shown in Fig.~\ref{fig:pes_dmrg} calculated by the QC-DMRG method using $M=16$ block 
states or alternatively by the TTNS approach 
with $z_i=2$ and $D=4$. We have confirmed that each state is a singlet with $S^2=0$. 

\begin{figure}[htb]
  \centerline{
    \includegraphics[scale=0.5]{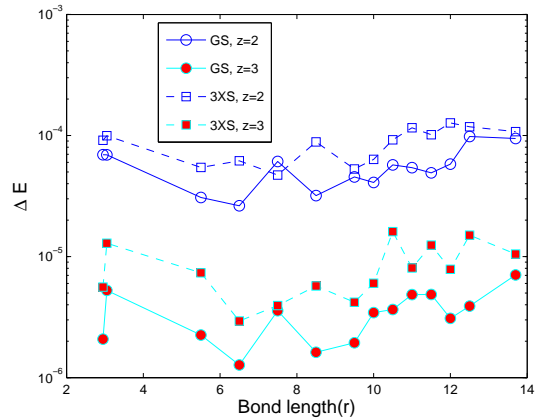}
        }
    \caption{(Color online) The relative error of the energy of the 
two lowest lying $^1\Sigma^+$ states as a function of bond length using the 
Tree-TNS with $z=2$ and $z=3$ with $D=4$. For $z=2$, the one-dimensional 
MPS-ansatz used in DMRG is recovered.
}
    \label{fig:pes_error_ttns}
\end{figure}
The relative error of the ground state ($\gamma=1$) and the third excited state ($\gamma=4$), 
$\Delta E_\gamma=|E_\gamma-E_{{\rm FCI}_\gamma}|/E_{{\rm FCI}_\gamma}$, are shown in Fig.~\ref{fig:pes_error_ttns}.
The open symbols stand for the MPS solution while the filled symbols indicate the TTNS result.
It is clear from the figure that the accuracy of the energy dropped for both states by at least
an order of magnitude. For the ground state close to the equilibrium bond length the change is almost two orders of magnitude.
It is important to emphasize again that all parameters of the calculations were kept fixed
except that we used the optimized one-dimensional MPS-like topologies or the two-dimensional optimized TTNS-like topologies.
The network topologies for both cases were optimized for each bond length
using the procedure outlined in Sec.~II.

\begin{figure}[htb]
  \centerline{
    \includegraphics[scale=0.5]{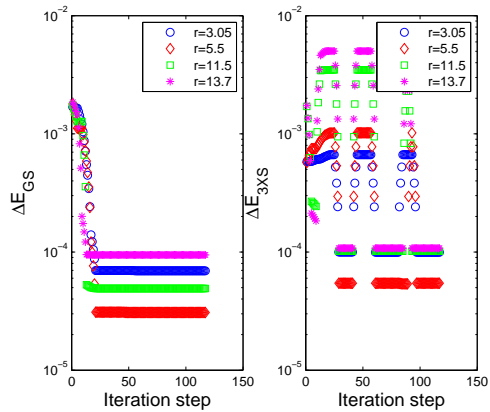}
        }
    \caption{(Color online) The relative error of the energy of the ground state and the third excited state
as a function of iteration step with $z_i=2$ and $D=4$ for a few selected bond lengths. 
}
    \label{fig:relerror_iter_mps}
\end{figure}

In the MPS based DMRG method the matrices of the one-dimensional tensor network are 
optimized iteratively by traversing through the network starting from the left boundary until the right 
boundary is reached. In the following steps the same procedure is repeated but in the reverse direction. This systematic optimization
procedure is called as sweeping. The relative error of the two $^1\Sigma^+$ states as a function of iteration steps
is shown in Fig.~\ref{fig:relerror_iter_mps} for the MPS case with $z_i=2$ and $D=4$.
It can be seen in the figure that the relative error of the ground state 
energy drops quickly as a function of iteration steps and it saturates after 
some 20 iterations. In contrast to this, the convergence of the third 
excited state is somewhat slower. In addition, the  
method lost the target state for certain interaction steps, i.e., 
for certain network configurations. This is due
to the very low bond dimension used in the test calculations.
When we used larger bond dimension or included point group symmetries 
this problem was fully eliminated.

\begin{figure}[htb]
  \centerline{
    \includegraphics[scale=0.4]{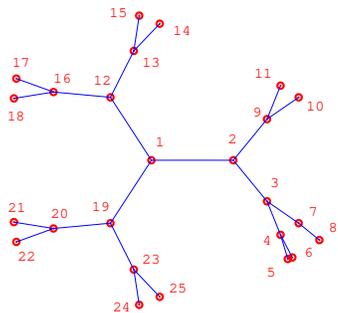}
        }
    \caption{(Color online) The figure shows how the tree network is traversed through in a full sweep. 
}
    \label{fig:ttns_sweeping}
\end{figure}

\begin{figure}[htb]
  \centerline{
    \includegraphics[scale=0.5]{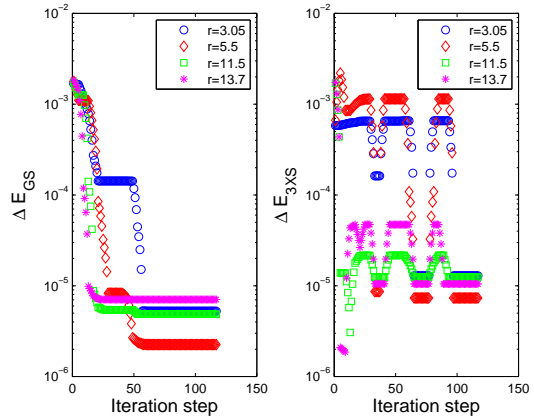}
        }
    \caption{(Color online) Similar to Fig.~\ref{fig:relerror_iter_mps} but for the TTNS method with $z=3$.  
}
    \label{fig:relerror_iter_ttns}
\end{figure}

In case of the tree-network, there is more freedom to choose the optimal sweeping procedure, i.e., 
to choose the optimal path through which the network is traversed. In the present work, we have 
swept through the network by going recursively back and forth through
 each branch. Therefore, according to the labeling of the orbitals on the lattice shown in 
Fig.~\ref{fig:ttns_sweeping} one sweep goes through the orbitals as
 1  2  3  4  5  4  6  4  3  7  8  7  3  2  9 10  9 11  9  2
 1 12 13 14 13 15 13 12 16 17 16 18 16 12  1 19 20 21 20 22 20 19 23 24 23 25 23 19.
The main advantage of this path is that highly entangled orbitals located close to the center of the network 
are optimized several times in a full sweep.
The relative error of the two $^1\Sigma^+$ states as a function of iteration steps
obtained by the TTNS method with $z_i=3$ and $D=4$ is shown in Fig.~\ref{fig:relerror_iter_ttns}. 
It can be seen that the relative error of the ground state energy
reached the saturation value of the MPS calculation (shown in 
Fig.~\ref{fig:relerror_iter_mps}) after a few iteration steps. 
However, unlike to the MPS result the error dropped further for subsequent
iteration steps until a much lower saturation value was reached.
The overall improvement compared to the MPS case was almost two orders 
of magnitudes. Similar improvement was observed for the excited state, although,
due to the low bond dimension the target state was lost again for 
certain interaction steps, i.e., for certain network configurations.

\subsection{Locating avoided crossing by entanglement}

\begin{figure}[htb]
        \centerline{
        \includegraphics[scale=0.5]{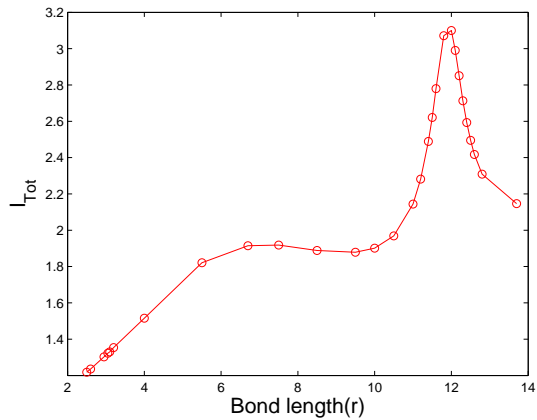}
        }
    \caption{(Color online) Total quantum information encoded in the wavefunction as a function bond length. The cusp-like structure 
             indicates the dramatic change in the wavefunction.}
    \label{fig:I_tot}
\end{figure}

Since the one- and two-orbital entropy functions were calculated to optimize network topologies
they can also be used to locate the avoided crossing. 
For problems 
in condensed matter physics the one- and two-orbital entropy functions and the block entropy are used to locate 
quantum phase transitions \cite{legeza2005-qpt,legeza2006-qpt}.
For translationally invariant systems the single-orbital entropy function is the same for all sites while
in a chemical system it is orbital dependent. Therefore, the behavior of $I_{\rm tot}$ 
as a function of bond length can be used
to detect and locate transition points where the wavefunction changes dramatically.  
In Fig.~\ref{fig:I_tot} $I_{\rm tot}$ is shown as a function of bond length. 
It can be seen 
that $I_{\rm tot}$ has a cusp like structure 
at $r=11.86$ indicating the position of the avoided crossing. 

\section{Conclusion}

The present paper has been devoted to the application of
the quantum-chemistry tree tensor network state
(QC-TTNS) method to calculate the potential energy
curve in the vicinity of the ionic-covalent avoided crossing in LiF.
We have discussed the main features of the most general version of the 
QC-TTNS algorithm in which the local properties of the tensors can be 
different for each orbital.  
The optimized tensor topologies, which reflect the structure 
of the entanglement bonds between the molecular orbitals, 
were determined by minimizing the entanglement length in the system 
as the bond length between the Li and F was stretched.
In order to compare tensor topology effects only, we have kept all parameters
of the algorithms fixed and used a very small bond dimension or
alternatively a very small number of block states.
By comparing the MPS(DMRG) and TTNS convergence properties we have demonstrated that the 
TTNS approach can converge to a significantly lower energy.  
Although the tensor contraction scales as $D^{z+1}$ the TTNS topology
offers a more optimal network structure since  
the relative error of the energy of the ground state as well as the excited states can be improved 
by an order of magnitude or more for the same value of $D$. 
This also indicates that the MPS result can be reproduced with a significantly lower bond dimension 
using the TTNS method.

Our QC-TTNS approach seems to be a promising direction reflected by the 
stability and fast convergence of the new method even for
systems in which the wavefunction character changes as
a function of bond length, especially in the region of an
avoided crossing. 
In contrast to the MPS case, however, the optimization of the network is a more complex task.
Extension of this work using more complex systems and orbital dependent 
coordination number is under progress.

\acknowledgments{
This research was supported by the 
European Research Area Chemistry (ERA-Chemistry) in part by   
the Hungarian Research Fund (OTKA) under Grant No. NN110360 and K100908,
the DFG SCHN 530/9-1 project under Grant No. 10041620 
and FWF-E1243-N19. V.M. and F.V. acknowledge support from the
SFB-ViCoM project.
\"O.L. acknowledges support from the Alexander von Humboldt foundation
and from ETH Z\"urich during his time as a visiting professor.
We are also grateful to Z. Rolik (Budapest) for providing
his FCI code based on the CI algorithm of Olsen et al.\cite{ofci}.
}


\begin{thebibliography}{10}

\bibitem{white}  
S.~R.~White, Phys.~Rev.~Lett. {\bf 69},  2863--2866  (1992).

\bibitem{white-qc}  
S.~R.~White and R.~L. Martin, J.\,Chem.\,Phys. {\bf 110},  4127--4130 (1999).

\bibitem{dmrg_lif}
\"O. Legeza, J. R\"oder and B. A. Hess, Molecular Physics {\bf 101}, 2019-2028 (2003).

\bibitem{chan-pes}
G.~K.-L. Chan, M. K\'{a}llay and G. J\"{u}rgen, J.\,Chem.\,Phys. {\bf 121}, 6110--6 (2004).

\bibitem{reiher-pes} 
G. Moritz, B. A. Hess, M. Reiher, J. Chem. Phys. {\bf 122}, 024107, (2005).

\bibitem{Kurashige2009}
Y. Kurashige and T. Yanai, J. Chem. Phys. {\bf 130}, 234114 (2009). 

\bibitem{boguslawski2013a} 
K.~Boguslawski, P.~Tecmer, G.~Barcza, \"O.~Legeza, and M.~Reiher, J.~Chem.~Theory Comp. {\bf 9} 2959 (2013).

\bibitem{legeza-rev} 
\"O.~Legeza, R.~Noack, J.~S{\'o}lyom, and L.~Tincani, in Computational Many-Particle Physics, eds. H.~Fehske,
R.~Schneider, and A.~Weisse {\bf 739}, 653--664 (2008).

\bibitem{reiher-rev} 
K.~H.~Marti and M.~Reiher, Z.~Phys.~Chem. {\bf 224},  583-599  (2010).

\bibitem{chan-rev} 
G.~K.-L.~Chan and S.~Sharma, Annu. Rev. Phys. Chem. {\bf 62},  465--481 (2011).

\bibitem{legeza03c} 
\"O. Legeza and J. S\'olyom, Phys. Rev. B {\bf 68}, 195116 (2003), 

\bibitem{legeza04a} 
\"O. Legeza and J. S\'olyom, Phys. Rev. B {\bf 70}, 205118 (2004).

\bibitem{rissler06} 
J. Rissler, R.M.Noack, and S.R. White, Chemical Physics, {\bf 323}, 519 (2006).

\bibitem{barcza2010a} 
G. Barcza, \"O. Legeza, K. H. Marti, and M. Reiher, Phys. Rev. A {\bf 83}, 012508 (2011).

\bibitem{boguslawski2012b} 
K.~Boguslawski, P.~Tecmer, \"O.~Legeza, and M.~Reiher, J.~Phys.~Chem.~Lett. {\bf 3}, 3129--3135 (2012).

\bibitem{yanai2013} 
Y. Kurashige, G.~K.-L. Chan and T. Yanai, Nature Chemistry, DOI: 10.1038/NCHEM.1677 (2013).

\bibitem{ostlund}
S. \"Ostlund and S. Rommer Phys. Rev. Lett. {\bf 75}, 3537 (1995). 

\bibitem{verstraeteciracmurg08}
F. Verstraete, J.I. Cirac, V. Murg, Adv. Phys. {\bf 57} (2), 143 (2008).

\bibitem{schollwock2011} U. Schollw\"ock, Ann. Phys. (NY) {\bf 326}, 96 (2011).

\bibitem{Haegeman2013} J. Haegeman, T. J. Osborne, and F. Verstraete,
arXiv:1305.1894 (2013).
 

\bibitem{verstraetecirac04}
F. Verstraete and J. I. Cirac, arXiv:cond-mat/0407066v1 (2004).

\bibitem{murgverstraete07}
V. Murg, F. Verstraete, and J. I. Cirac, Phys. Rev. A, {\bf 75}, 033605 (2007).

\bibitem{murgverstraete08}
V. Murg, F. Verstraete, and J. I. Cirac, Phys. Rev. B, {\bf 79}, 195119 (2009).


\bibitem{vidal06}
G. Vidal, Phys. Rev. Lett., 101, 110501 (2008)

\bibitem{changlani09}
H. J. Changlani, J. M. Kinder, C. J. Umrigar, and G. K.-L. Chan, Phys. Rev. B, {\bf 80}, 245116 (2009). 

\bibitem{marti10}
K. H. Marti, B. Bauer, M. Reiher, M. Troyer, and F. Verstraete, New J. Phys. {\bf 12} 103008 (2010).

\bibitem{marti-rev}
K. H. Marti, M. Reiher, Phys. Chem. Chem. Phys. {\bf 13} 6750-6759 (2011).


\bibitem{schneider-rev}
\"O. Legeza, T. R\"ohwedder, and R. Schneider: Numerical approaches for high-dimensional PDE's for quantum chemistry, 
in Encyclopedia of Applied and Computational Mathematics,Editor-in-chief: Engquist, Bj\"orn ; Chan, T.; Cook, W.J.; 
Hairer, E.; Hastad, J.; Iserles, A.; Langtangen, H.P.; Le Bris, C.; Lions, P.L.; Lubich, C.; Majda, 
A.J.; McLaughlin, J.; Nieminen, R.M.; ODEN, J.; Souganidis, P.; Tveito, A. (Eds.) Springer 2013 ISBN 978-3-540-70530-7 

\bibitem{schneider-rev2013}
\"O. Legeza, T. Rohwedder, R. Schneider, and Sz. Szalay, arXiv:1310.2736 (2013).

\bibitem{hackbusch} W. Hackbusch, \emph{Tensor Spaces and Numerical Tensor Calculus}, SSCM Vol. {\bf 42}, Springer, 2012.

\bibitem{orus-rev} R. Orus, arXiv:1306.2164 (2013). 

\bibitem{murg2010-tree}
V. Murg, F. Verstraete, \"O. Legeza, and R. M. Noack, Phys. Rev. B {\bf 82}, 205105 (2010).

\bibitem{chan2013-tree}
N.~Nakatani and G.~K.-L. Chan, J. Chem. Phys. {\bf 138}, 134113 (2013).


\bibitem{shi2006}
Y.~Shi, L.~Duan, G.~Vidal, Phys. Rev.  {\bf 74}, 02232 (2006).


\bibitem{Tagliacozzo2007} L. Tagliacozzo, G. Evenbly, and G. Vidal, Phys. Rev. B {\bf 80}, 235127 (2009).

\bibitem{Corboz2009} P. Corboz and G. Vidal, Phys. Rev. B {\bf 80}, 165129 (2009).

\bibitem{Corboz2010} P. Corboz, G. Evenbly, F. Verstraete, and G. Vidal, Phys. Rev. A {\bf 81}, 010303(R) (2010).

\bibitem{Lubich} C.~Lubich, T.~Rohwedder, R.~Schneider, B.~Vandereycken,
SIAM J. Matrix Anal. Appl. {\bf 34} 470-494 (2103)


\bibitem{legeza2003a}
\"O. Legeza, J. R\"oder, and B. A. Hess, Phys. Rev. B {\bf 67}, 125114 (2003).



\bibitem{mcculloh}
I.P. Mcculloch, J. Stat. Mech., P10014 (2007).

\bibitem{Toth2008}
A. I. T\'oth, C. P. Moca, \"O. Legeza, and G. Zar\'and, Phys. Rev. B {\bf 78}, 24510 (2008).

\bibitem{Zgid2008}
D. Zgid and M. Nooijen, J. Chem. Phys. {\bf 128}, 014107 (2008).

\bibitem{Sharma2012}
S. Sharma and G.~K.-L. Chan, J. Chem. Phys. {\bf 136}, 014107 (2012).

\bibitem{Wouters2013}
S. Wouters, W. Poelmans, P. W. Ayers, and D. {Van Neck}, 
arXiv:1312.2415 (2013).

\bibitem{schollwock2005} 
U. Schollw\"ock, Rev. Mod. Phys. {\bf 77}, 259 (2005).

\bibitem{fci_lif}
C. W. Bauschlicher and S. R. Langhoff J. Chem. Phys., {\bf 89}, 4246-425 (1988).

\bibitem{gamess}
M.S.Gordon and M.W.Schmidt, "{\rm in:} Theory and Applications of Computational Chemistry: the first forty years", 
1167-1189, Elsevier, Amsterdam, (2005).


\bibitem{ofci}
J. Olsen and B. O. Roos and P. Jorgensen and H. J. Aa. Jensen, J. Chem. Phys., {\bf 89}, 2186 (1988).

\bibitem{legeza2005-basis}
\"O. Legeza, F. Gebhard, and J. Rissler, Phys. Rev. B {\bf 74}, 195112 (2006).

\bibitem{fiedler1}
M. Fiedler, Czech. Math. Journal {\bf 23}, 298 (1973).

\bibitem{fiedler2}
M. Fiedler, Czech. Math. Journal {\bf 25}, 619 (1975)

\bibitem{legeza2005-qpt}
\"O. Legeza and J. S\'olyom, Phys. Rev. Lett. {\bf 96}, 116401 (2006). 

\bibitem{legeza2006-qpt}
\"O. Legeza, J. S\'olyom, L. Tincani, and R. M. Noack, Phys. Rev. Lett. {\bf 99}, 087203 (2007).

\bibitem{boguslawski2013b}
P. Tecmer, K. Boguslawski, O. Legeza, and M. Reiher, Phys. Chem. Chem. Phys. {\bf 16}, 719 (2014). 
 
\bibitem{berkman93}
O. Berkman and U. Vishkin, SIAM J. Comp. 22 (2): 221-242 (1993).

\end{thebibliography}
\end{document}